\documentclass[prd,preprint,tightenlines,superscriptaddress,floatfix,showpacs,
preprintnumbers,nofootinbib,eqsecnum]{revtex4}

 \usepackage[dvips,final]{graphicx}
     \usepackage{epsfig}
      \usepackage{bm}

\begin{document}

\thispagestyle{empty} \preprint{\hbox{}} \vspace*{-10mm}

\title{Studying the Interplay of Strong and Electromagnetic Forces in 
Heavy Ion Collisions with NICA\\
Research Proposal}

\author{A.~Rybicki}
\email{andrzej.rybicki@ifj.edu.pl}

\affiliation{H.Niewodnicza\'nski Institute of Nuclear Physics, Polish Academy of Sciences, Radzikowskiego 152, 31-342 Krak\'ow, Poland}

\author{A.~Szczurek}
\email{antoni.szczurek@ifj.edu.pl}

\affiliation{H.Niewodnicza\'nski Institute of Nuclear Physics, Polish Academy of Sciences, Radzikowskiego 152, 31-342 Krak\'ow, Poland}
\affiliation{University of Rzesz\'ow, Rejtana 16, 35-959 Rzesz\'ow, 
Poland}

\author{M.~K{\l}usek-Gawenda}

\affiliation{H.Niewodnicza\'nski Institute of Nuclear Physics, Polish Academy of Sciences, Radzikowskiego 152, 31-342 Krak\'ow, Poland}

\author{I.~Sputowska}

\affiliation{H.Niewodnicza\'nski Institute of Nuclear Physics, Polish Academy of Sciences, Radzikowskiego 152, 31-342 Krak\'ow, Poland}

\date{\today}

\begin{abstract}

 In the following we stress the advantages of the NICA research programme in the context of studying the spectator-induced electromagnetic phenomena present in proton-nucleus and heavy ion collisions. We point at the specific interest of using these phenomena as a new, independent source of information on the space-time evolution of the reaction and of the non-perturbative process of particle production. We propose an extended series of measurements of well-defined observables to be performed in different types of nuclear reactions and in the whole range of collision energies available to NICA. We expect these measurements to bring very valuable new insight into the mechanism of non-perturbative strong interactions, complementary to the studies made at the SPS at CERN, RHIC at BNL, and the LHC. 
\end{abstract}


\begin{figure}[b]             
\centering
\includegraphics[width=0.75\textwidth]{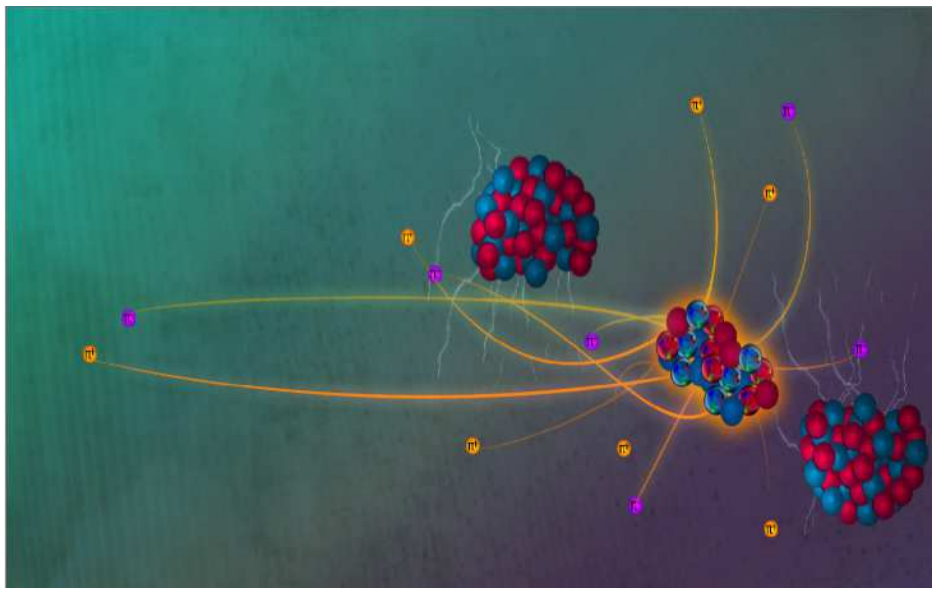}
   \caption{The general idea of the spectator-induced electromagnetic effect discussed in this paper. 
The figure is redrawn from~\cite{sputowskameson2012}.}
\label{fig:figzero}
\end{figure}

\maketitle

\section{Introduction}
\label{secone}

The specific features of the NICA (Nuclotron based Ion Collider 
fAcility) research programme~\cite{trubnikov-nica}, and in particular 
(a) the broad range of reactions planned to be studied, (b) the 
relatively modest collision energy $\sqrt{s_{NN}}$, placing the whole 
programme close to that of the CERN SPS, (c) the possibility to measure 
reactions at different energies, and (d) the elasticity of taking 
measurements both in fixed target and collider modes, make it well 
suitable for studying the interplay of strong and electromagnetic 
interactions in nuclear collisions. What we specifically address in the 
present proposal is the {\em electromagnetic interaction between charged 
particles produced in the collision and the nuclear remnant that does 
not participate directly in the reaction} (the „spectator system”). This 
latter phenomenon is of particular interest because, as discussed in our 
earlier works~\cite{twospec}, {\em it provides independent information 
on the space time-evolution of the reaction}: the space-time evolution 
of the particle production process, the fragmentation (break-up) of the 
spectator system, and the interplay between the two. It is to be noted 
that with the well-known exception of HBT measurements~\cite{baym}, the 
experimental programmes at the SPS, RHIC and LHC provide direct 
information essentially only on the final (or near-to-final) state 
particles in {\em momentum space} ($p_x$, $p_y$, $p_z$). Much less is 
possible as far as providing information on the evolution of the 
reaction in {\em position space} ($x$,$y$,$z$), which on the other hand 
is extremely important in view of our understanding of the heavy ion 
reaction. Here NICA could strongly contribute to the overall knowledge 
in the whole heavy ion field, with very little competition from existing 
experiments. The corresponding measurements would in particular include 
{\em particle spectra, charged particle ratios} ($\pi^+/\pi^-$, 
$K^+/K^-$, etc.) and {\em directed flow}.

This paper is organized as follows. In Section~\ref{sectwo}, we discuss 
the principal features of the spectator-induced electromagnetic effect, 
with particular emphasis on its importance as a new source of 
information on the mechanism of the nuclear reaction. In 
Section~\ref{secthree}, we 
define the possible future contribution of NICA. In 
Section~\ref{secfour}, we 
shortly address the subject of competition from other experiments. We 
present our conclusions in Section~\ref{secfive}.

\section{What do we know about the spectator-induced electromagnetic effect?}
\label{sectwo}

It is not surprising that various kinds of electromagnetic interactions 
in nucleus-nucleus collisions were studied in the past (a partial 
overview can be found in~\cite{bartke,habrybicki}). The problem of 
Coulomb corrections to HBT measurements belongs in fact to the 
“standard” in the analysis of heavy ion experimental data. Numerous 
works exist also on the influence of the electromagnetic field on the 
spectra of particles produced at {\em mid-rapidity} (that is, in the 
vicinity of $x_F=0$) in central heavy ion reactions\footnote{We always 
define the Feynman $x_F = \frac{p_L}{p_L(max)}$ and rapidity $y = 
\frac{1}{2} \ln\left(\frac{E+p_L}{E-p_L}\right)$ in the nucleon-nucleon 
c.m.s.}~\cite{e802}. However, these concentrate on the electromagnetic 
field induced by the presence of initial charge in the “participant 
zone”, that is, the charge of the {\em participating 
nucleons}\footnote{Note:
 the two electromagnetic effects discussed above, namely that induced by 
the participant and by the spectator charge, should be clearly 
differentiated as they have a different distribution over phase-space 
and a different centrality dependence. The participant charge will 
mostly influence the region closer to $x_F=0$ in central nucleus-nucleus 
collisions, see above, while the spectator charge will produce the 
largest effect at higher values of $x_F$ and in peripheral collisions.}. 
On the other hand, it seems that the theoretical and experimental 
analyses of the electromagnetic interactions between {\em nuclear 
remnants} and produced particles were performed mostly at lower 
energies. Very sizeable electromagnetically-induced distortions were 
observed there~\cite{beneson}. An important, much more recent result, 
was reported in nuclear collisions at several GeV/nucleon, where the 
non-relativistic approach to the Coulomb field brought information on 
the space-time evolution of the process of nuclear 
fragmentation~\cite{karnaukhov}. In the energy regime of the CERN SPS 
and above, up to now our main range of interest, we were aware of only 
one earlier experimental measurement~\cite{na52ambrosini99}. The latter 
was unfortunately restricted to a very narrow acceptance range (forward 
angles i.e. $p_T\approx 0$), and to an extremely small number of data 
points (between two and four, depending on collision centrality) which 
limited its scientific usefulness.

\begin{figure}[t]             
\centering

\includegraphics[width=0.85\textwidth]{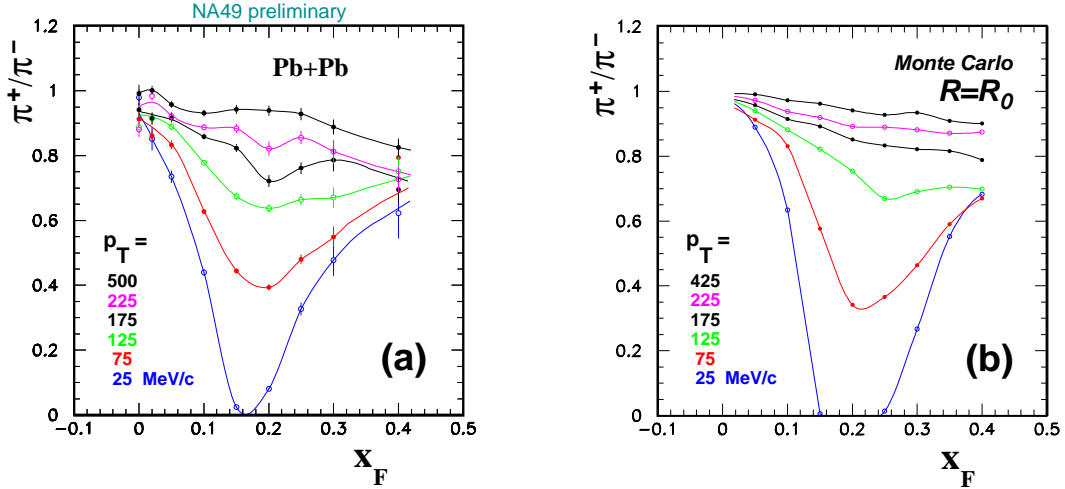}
   \caption{
 Electromagnetic distortion of charged pion ratios in the 
projectile hemisphere of the peripheral Pb+Pb reaction. (a) The 
$\pi^+/\pi^-$ ratios measured by the NA49 experiment at the CERN SPS, 
drawn as a function of $x_F$ at fixed values of $p_T$ (listed from top 
to bottom curve); (b) Result of the Monte Carlo simulation described in 
the text. The figure comes from~\cite{rybickisqm2011}.}
 \label{fig:figone}
\end{figure}

\begin{figure}[t]             
\centering
\includegraphics[width=0.75\textwidth]{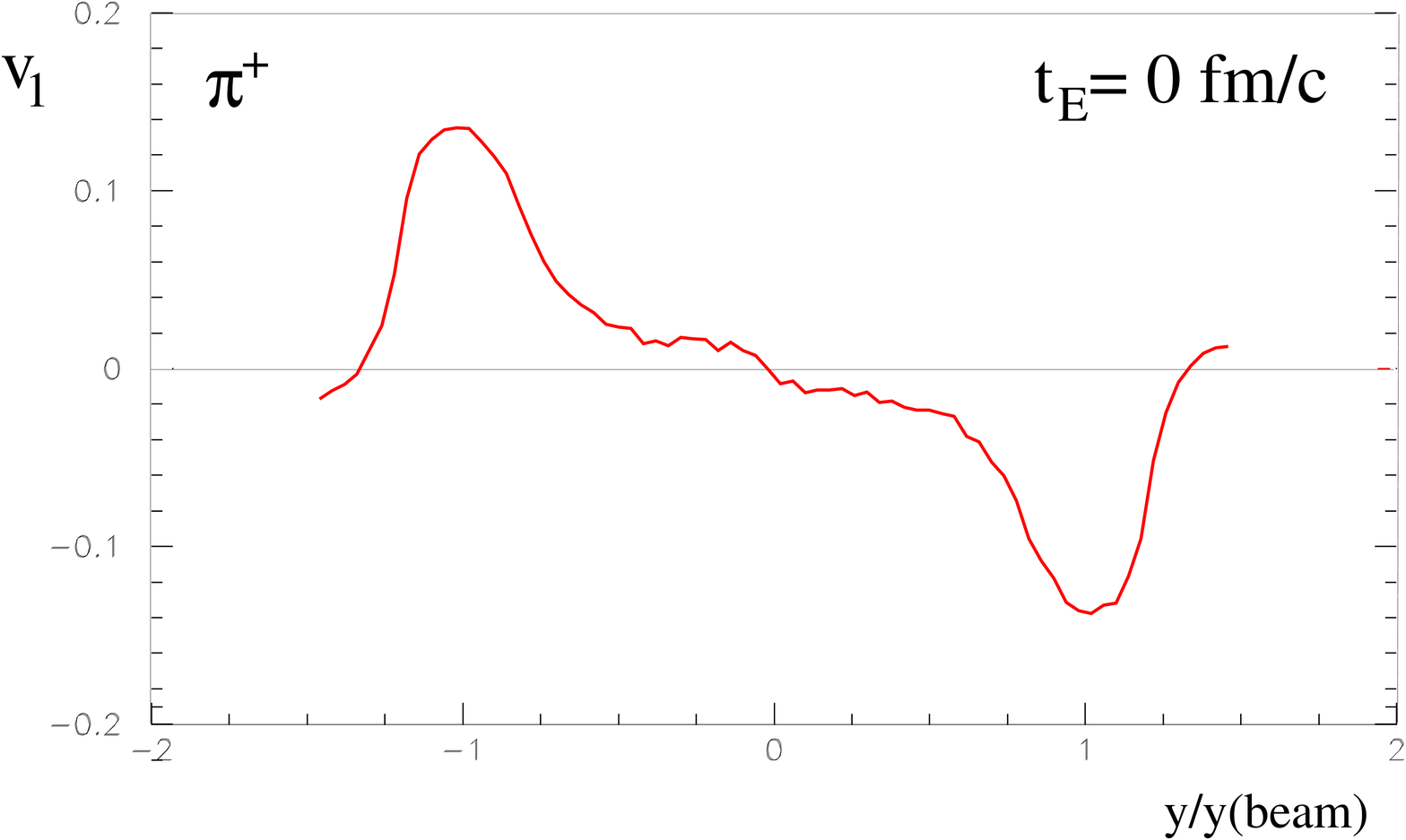}
   \caption{Dependence of the electromagnetically-induced directed flow 
of positively charged pions on the scaled pion rapidity $y/y(beam)$. The 
simulation assumes the pion emission time $t_E$ equal to zero. Directed 
flow is integrated over $p_T$ from 0 to 1 GeV/c.}
 \label{fig:figthree}
\end{figure}

For this reason, we performed a series of experimental and theoretical 
studies of the influence which the spectator charge exerts on charged 
pion and charged kaon spectra\footnote{All the experimental results were 
obtained within the framework of the NA49 experiment at 
the CERN 
SPS (see~\cite{na49nim} for a detailed description of the NA49 
detector).}~\cite{twospec,posrybicki,szczurekqm06,rybickiapb2011z,rybickisqm2011,sputowskameson2012}. 
Out of these, a consistent picture emerges which can be summarized as 
follows:

 \begin{enumerate}

 \item[\bf{1.)}]
 The presence of the spectator-induced electromagnetic field brings a 
very sizeable distortion to $\pi^+/\pi^-$ ratios observed in the final 
state of “peripheral” (large impact parameter) Pb+Pb reactions measured 
at a beam energy of 158 GeV/nucleon ($\sqrt{s_{NN}}=17.3$~GeV). This is 
shown in 
Fig.~\ref{fig:figone}(a), where the $x_F$-dependence of $\pi^+/\pi^-$ 
ratios is drawn in 
the projectile hemisphere of the reaction, for fixed values of pion 
transverse momentum $p_T$. This effect is so strong that the 
$\pi^+/\pi^-$ 
ratio goes close to zero in the vicinity of $x_F=0.15$, violating 
isospin 
symmetry and thus unequivocally confirming the electromagnetic origin of 
the whole phenomenon. Note that the latter value of $x_F=0.15=m_\pi/m_N$ 
corresponds, at low transverse momenta, to pions moving at the same 
velocity as the spectator system, thus confirming that electromagnetic 
repulsion (attraction) of positive (negative) pions from positively 
charged spectator protons is indeed at the cause of this behaviour.

\begin{figure}[t]             
\centering
\includegraphics[width=0.75\textwidth]{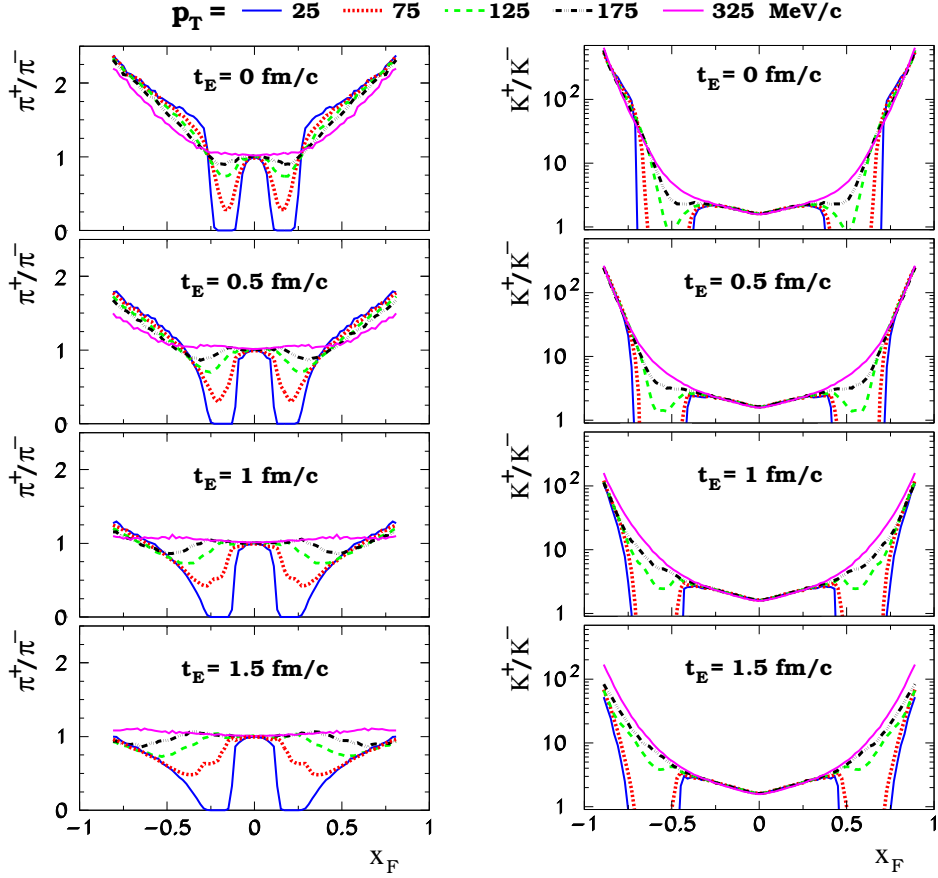}
   \caption{Dependence of the electromagnetic distortion of 
$\pi^+/\pi^-$ (left) and $K^+/K^-$ (right) ratios, for particles 
produced 
in 
peripheral Pb+Pb collisions. The different panels correspond to 
different pion and kaon emission times $t_E$. The figure comes 
from~\cite{rybickisqm2011}.}
 \label{fig:figtwo}
\end{figure}

 \item[\bf{2.)}]
 Our simple Monte Carlo model of the electromagnetic 
interaction~\cite{twospec,posrybicki} brings a very reasonable 
description of the main features of this effect as shown in 
Fig.~\ref{fig:figone}(b). 
The unique region where a more significant disagreement between data and 
model can be seen ($x_F\approx 0.2$, low $p_T$) has been identified as 
due 
to the 
process of {\em nuclear fragmentation (break-up) of the spectator 
system}; a discussion of this subject can be found 
in~\cite{rybickiapb2011z}.

 \item[\bf{3.)}]
 A very similar electromagnetic effect is also 
present in high energy 
collisions of lead ions with lighter nuclei which we recently 
investigated~\cite{sputowskameson2012}.

 \item[\bf{4.)}]
 The electromagnetic distortion observed in 
Fig.~\ref{fig:figone}(a) {\em depends on 
the specific space-time scenario imposed on pion emission}. This is 
illustrated in Fig.~\ref{fig:figtwo} (left) where the results of our 
model calculations 
are drawn in the full range of $x_F$, $−1<x_F<1$, for different values 
assumed for the time of pion emission\footnote{Note: we define the time 
of pion emission with respect to the moment of closest approach of the 
two colliding nuclei~\cite{twospec}.} $t_E$. The characteristic 
distortion 
pattern imposed by the two spectator systems at positive and negative 
$x_F$ 
appears clearly sensitive to $t_E$ (with typically lower $\pi^+/\pi^-$ 
ratios obtained for larger $t_E$ at higher values of $x_F$). {\em This 
implies 
that the electromagnetic effect provides independent information on the 
evolution of the non-perturbative process of pion production in space 
and time}~\cite{twospec,posrybicki}.

 \item[\bf{5.)}]
 The electromagnetic distortion of charged kaon 
($K^+/K^-$) ratios, 
Fig.~\ref{fig:figtwo} (right), exhibits basic qualitative similarities 
to 
the effect 
seen for pions, however, with pronounced differences on the 
quantitative 
level. The position of the deep “valley” in the ratio is displaced 
towards higher values of $x_F$, and the region of highest sensitivity to 
the kaon emission time is moved towards very high $x_F$ ($x_F>0.75$).

 \item[\bf{6.)}]
 Finally, as apparent in Fig.~\ref{fig:figthree}, the 
spectator-induced 
electromagnetic force exerts also {\em a noticeable influence on pion 
directed flow, $v_1$}. Our very recent Monte Carlo calculation
(not yet published) 
predicts a well-defined pattern in the rapidity dependence 
of the {\em electromagnetically-induced} directed flow of positive 
pions, with a large peak in the vicinity of projectile and target 
rapidities. As such, directed flow of charged pions ($\pi^+$, $\pi^-$) 
appears as another observable where, also through electromagnetic 
effects, new information on the space-time evolution of the reaction can 
become available.

\end{enumerate}

The above description is, of course, highly simplified (the interested 
reader is invited to visit the enclosed bibliography for a more detailed 
account), but nevertheless constitutes a good starting point for the 
present proposal. As it comes from the considerations above, and as has 
also been suggested to us in numerous discussions with experts, the 
spectator-induced electromagnetic effects discussed here have very 
specific characteristics which make them attractive for numerous future 
studies:

 \begin{enumerate}
 \item[-]
 they can be large in specific regions of phase space;

 \item[-]
 they can bring large distortions to various collision 
characteristics 
observed in the final state (like charged particle ratios or directed 
flow) and at the same time, they can provide information about the 
intrinsic space-time scenario of particle emission (formation times, 
parton fragmentation, resonance decays, hydrodynamics, etc).
 \end{enumerate}

As such, in principle they should be studied in any possible reaction at 
any possible energy, whenever this is possible. This is particularly 
important in the present situation where the available experimental 
information remains very limited. In the high energy regime (SPS 
energies and above), no experimental data set other than what was 
discussed above is known to us. No possibility of obtaining such 
information at RHIC or LHC is apparent to us due to strict experimental 
limitations.  For this reason, we see here a good scientific prospect 
for NICA.

\section{Possible measurements at NICA}
\label{secthree}

The principal detector requirements needed in order to perform 
measurements of the spectator-induced electromagnetic effect are 
particle momentum vector reconstruction and particle identification 
capabilities (including in particular charge differentiation), as well 
as a relatively wide acceptance coverage defined in terms of 
longitudinal and transverse momenta. Once these conditions are 
fulfilled, measurements of particle spectra (preferably double 
differential spectra of the type $\frac{d^2n}{dydp_T}$) become 
accessible. 
Measurements of directed flow (as well as possibly higher harmonics) are 
characterized by additional requirements well known to the community.

In the range of collision energies specified in~\cite{trubnikov-nica}, 
the NICA/MPD apparatus looks promising in view of the requirements 
specified above. With the extremely broad spectrum of reactions planned 
to be analysed, including in particular proton-nucleus and 
nucleus-nucleus collisions with an impressive versatility of projectiles 
and targets, we propose a {\em detailed, systematic experimental study 
of charged particle spectra and directed flow with a special emphasis on 
charge asymmetries induced by electromagnetic interactions}. Such a very 
broad study, taking full benefit of the possibility of cross-comparisons 
between different reactions (in particular also proton-nucleus 
collisions), of switching between fixed target and collider modes with 
different ranges of effective acceptance in both cases, and of 
comparisons of different collision energies, would in our view provide 
new information on the {\em space-time evolution of very different 
aspects of nuclear reactions}. In particular, these would be:

 \begin{enumerate}
 \item[-]
 the centrality and system-size dependence of particle 
production phenomena;

 \item[-]
 the interplay between particle production and spectator 
fragmentation;

 \item[-]
 azimuthal anisotropies and flow; 

 \item[-]
 the energy dependence of particle production.

\end{enumerate}

All of the above are clearly basic, essential issues in 
heavy ion 
collision physics. With the advent of the NICA research programme, 
these could be analysed also with full use of the specific 
electromagnetic interaction as discussed in the present proposal.

\section{Competition from existing experiments}
\label{secfour}

As already specified above, we consider the competition from existing 
experiments as relatively weak with respect to the possibilities 
offered by NICA and its experimental community. We see at present no 
RHIC nor LHC experiment to be able to provide reliable measurements of 
the 
phenomena discussed here. What remains is clearly the SPS energy range 
where, as evident from the results presented in Fig.~\ref{fig:figone} 
above, new 
experimental analyses can be performed on the basis of data from the 
NA49 experiment~\cite{na49nim} or from its extension, 
NA61/SHINE~\cite{na61}. This could possibly make these experiments 
complementary to NICA/MPD, but with specific limitations which would 
require a more in-depth study.

\section{Conclusions}
\label{secfive}

The interplay of strong and electromagnetic interactions in nuclear 
collisions, including in particular the spectator-induced 
electromagnetic distortion of charged particle spectra and azimuthal 
anisotropies, constitutes a new, and in our opinion, very promising 
source of information on the space-time evolution of various processes 
present in the heavy ion reaction. The range of collision energies 
considered for NICA, together with its experimental elasticity in terms 
of data taking mode and of the versatile choice of interacting 
projectiles and targets, makes it in our opinion well suitable to bring 
a very valuable contribution to studies of this phenomenon.

\vspace{1cm}

{\bf Acknowledgments}\\

This work was supported by the Polish National Science Centre 
(on the basis of decision no. DEC-2011/03/B/ST2/02634).


\end{document}